Comments on: "Increasing destructiveness of tropical cyclones over the past 30 years" by Kerry Emanuel, Nature, 31 July 2005, Vol. 436, pp. 686-688

William M. Gray

Department of Atmospheric Science

Colorado State University

Fort Collins, CO  80523   USA

## ABSTRACT

**The near universal references to the above paper by most of the major US media outlets and blogs since Katrina and Rita made US landfall requires a response from a few of us who study hurricanes.   Having been involved with hurricane research and forecasting for nearly 50 years, I feel I have an obligation to offer comments on this paper's findings which, in my view, are not valid.**

**This paper concludes that global tropical cyclone net power dissipation [(or friction times wind) - taken to be proportional to the sum of each cyclone's individual 6-hour track period maximum wind speed cubed (~ $V_{max}^3$)] has undergone large (more than doubled) increases over the last 30 years. The author associates these frictional energy dissipation increases with rising mean sea surface temperatures (SSTs) and implies that these SST increases may, in part, be related to human activity.  If true, this is a very important finding that has great relevance as regards to the globe's future climate and future hurricane**

**destruction. But, the author's 'apparent' blockbuster results and his interpretation of his calculations are not realistic.**

## DISCUSSION

This paper's conclusions are partially based on an analysis of poorly documented and biased tropical cyclone (TC) maximum wind ($V_{max}$) data from the Northwest (NW) Pacific. This study presents a sum of each tropical cyclone's maximum winds - a very hard measurement to realistically obtain over a 3-5 decade period of changing maximum wind measurement techniques. The author then cubes these maximum wind speeds ($V_{max}^3$) of questionable accuracy. Such a cubing process greatly exaggerates any errors in the original $V_{max}$ estimate. For instance, a TC with a 40 ms$^{-1}$ maximum wind speed and a plus/minus 10% margin of error (44 m/s versus 36 m/s) in $V_{max}$ wind leads to a 80 percent high vs. low difference in $V_{max}^3$.

## DETERMINATION OF HURRICANE MAXIMUM INTENSITY

There always has been, and there probably always will be, problems in assigning a representative maximum surface wind to a hurricane. As technology advanced from conditions during the mid-20$^{th}$ century to the present, the methods of determining a hurricane's maximum winds underwent change. These changes have introduced difficulties in realistically comparing long period tropical cyclone maximum wind records.

With the availability of new aircraft-deployed inertial dropwindsondes and the new step-frequency surface wind measurement instruments, it is being found that sustained surface winds in a hurricane are often stronger than previously estimated. The correction factor for maximum wind values extrapolated downward from aircraft altitude has changed in the Atlantic (Franklin et al. 2003). Atlantic hurricane category numbers have and likely will continue to undergo upward increases due to the continuing improvement of technology. Technology corrections will have to be made when $V_{max}$ comparisons are made of long period records. Similar but different technical correction needs are required in the Northwest (NW) Pacific.

The author has utilized the official best track NW Pacific tropical cyclone (TC) $V_{max}$ values over the last 50 years. He has made corrections to some of the $V_{max}$ values and has accepted others as realistic. His analysis has many problems. It is known that there are long-period systematic biases in the NW Pacific maximum wind ($V_{max}$) values because of the methodological changes in how the NW Pacific maximum wind estimates have been made. The author tried to account for these measurement technique differences, but his method of doing so was wrong. It was his faulty corrections for the long term biases in the NW Pacific data set which led to his invalid calculations that TC energy dissipation had shown large increases in recent years.

Prior to 1973, maximum winds ($V_{max}$) in the NW Pacific could not be reliably measured. They were estimated from aircraft central pressure and flight crew visual estimations of wave-spray characteristics. Crews had a tendency to exaggerate their maximum wind estimates. A guideline to determining maximum winds ($V_{max}$) from measured central pressure in general use at the Joint-Typhoon Warning Center (JTWC) until the early 1970s was $V_{max} = 16(1010 - Pc)^{0.5}$, where Pc is central pressure in millibars and $V_{max}$ is maximum wind in knots.

To correct what appeared, at the time to be overestimations of $V_{max}$, Atkinson and Holliday (1977), while engaged in forecasting duties at JTWC, devised a new maximum wind-central pressure relationship scheme which was used as a guide for later $V_{max}$ determination. This wind-pressure relationship was applied between 1973-1986. This new formula specified $V_{max} = 6.7 (1010 - Pc)^{0.644}$. These newer Atkinson-Holliday (AH) estimates gave central pressure-determined maximum wind values which are now known to be too low. For a typhoon with a MSLP of 950 mb, the $V_{max}$ difference by these two techniques is 32 percent lower with the Atkinson-Holliday (AH) scheme than the pre-1973 scheme. This would lead to a 232 percent higher value of $V_{max}^3$ with the earlier pre-1973 method than with the later (1973-86) AH method.

The AH scheme substantially underestimated $V_{max}$ for all typhoons. This resulted in a significant underestimate of Northwest Pacific cyclone $V_{max}$ values between 1973-1986. For instance, in the 6-year period prior to 1973 the average seasonal number of typhoons with maximum winds above 100 and 130 knots was over twice as much as during the average 6-year period of 1973-1978. The average annual number of super typhoons in the West Pacific between 1950-1972 and 1987-2004 was 5.3 and 4.9. In the AH period of 1973-86, it was only 2.3. Yet West Pacific typhoon activity in the 1973-86 periods showed no apparent differences from earlier or later period activity. The AH scheme was discontinued after 1986 when aircraft reconnaissance in the NW Pacific was terminated, and central pressures could no longer be directly measured. $V_{max}$ values have since been obtained solely from satellite. The Dvorak satellite TC intensity scheme (1975, 1984) is known to give systematic higher $V_{max}$ estimates than the AH scheme. Knaff and Zehr's (2005) recent analysis shows that the Dvorak satellite (1975, 1984) scheme for the estimation of $V_{max}$ (used in the Pacific since 1987) gives, on average, about 7.5 m/s higher $V_{max}$ value than the AH scheme for all wind speed categories. There is no question that the Dvorak scheme is superior to the AH scheme. The differences between the Dvorak and the AH schemes causes large differences in $V_{max}^3$. For 7.5 m/s wind differences of 32.5 m/s for Dvorak versus 25 m/s for AH, the cubed ratio of the Dvorak to the AH maximum wind speed is 2.2. For higher $V_{max}$ values (say 57.5 vs. 50 m/s) this ratio is 1.5. There are many more TC time periods in Emanuel's analysis of the lower $V_{max}$ values when this ratio of the cubed $V_{max}$ of the satellite to the AH is close to 2 to 1.

Most of the large increase in Emanuel's NW Pacific TC energy dissipation calculations from the early 1970s to the early 2000s can be explained by the cubed differences of the $V_{max}$ estimates between the more recent (since 1987) Dvorak pure satellite scheme for $V_{max}$ and the earlier (1973-1986) period when the AH aircraft scheme was used. Gray et al. (1991) and Martin and Gray (1993) have discussed the many complications that arise for $V_{max}$ determination from utilization of satellite-only versus aircraft-only measurements and when both measurements are available.

The rest of Emanuel's last 30-year increase in NW Pacific tropical cyclone frictional energy dissipation can be explained by a variety of other decadal changes unique to the NW Pacific including increased El Niño activity in recent decades. Northwest Pacific tropical cyclones form further eastward in El Niño years and have longer over-ocean tracks. El Niño activity has been more frequent during the last three decades. When El Niño activity is reduced in future years, NW Pacific track length will likely be reduced as will Emanuel's Power Dissipation Index (PDI).

Another major problem with Emanuel's results is his assumption that a measurement of $V_{max}$ is directly related to the net frictional energy dissipation of the cyclone's entire broad scale circulation. A cyclone's outer radius area is vastly larger than its inner core of high wind speed and can bring about more net wind energy dissipation than the inner-core. Emanuel's use of $V_{max}^3$ for an estimation of TC net broad scale energy dissipation is not a satisfactory assumption.

Merrill (1982) and Cocks and Gray (2002) found that a cyclone's size (as measured by the outer closed isobar or by the mean radius of 15 or 25 m/s winds) is a much better representation of the cyclone's net tangential momentum than its inner core intensity expressed by $V_{max}$. Many years of studying a tropical cyclone's net area of cyclonic wind (as determined by its mean outer closed isobar) and relating it to its $V_{max}$ in both the NW Pacific and the Atlantic showed that these two factors were correlated at about 0.3, explaining only about 10 percent of the variance between $V_{max}$ and the cyclone's outer area of tangential wind. There are also life cycle differences in tropical cyclones. As cyclones intensify (in their early stages) their ratio of $V_{max}$ to outer wind strength ratio is significantly higher than when they are filling at higher latitudes (Weatherford and Gray I and II, 1987).

In the Atlantic, $V_{max}$ has been more reliably measured with aircraft. But increases in $V_{max}^3$ and longer cyclone tracks have occurred only in the last 11 years (1995-2005) and not over the last 30 years as this study implies. Most of Emanuel's 30-year increase in wind energy dissipation occurs only from the combination of the Atlantic and Pacific data sets where the Pacific has nearly three times as many tropical cyclones as the Atlantic.

There was a lower Atlantic net $V_{max}^3$ between 1975 and 1994, at a time that global SST's were rising.  Much higher $V_{max}^3$ values occurred in the 1950s and 1960s when global mean SST values were undergoing a modest decrease.  It is only in the last decade that there have been increases in Atlantic $V_{max}$ and track length.  The last-decade Atlantic increases in hurricane activity can be directly associated with the large increase in the strength of the Atlantic Ocean Thermohaline circulation (THC) – the so called positive phase of the Atlantic multi-decadal oscillation (AMO).  Multi-decadal variations in Atlantic sea surface temperatures, associated with multi-decadal oscillations in Atlantic hurricane activity, have also been documented in Greenland ice-core temperature records going back thousands of years.  The increase in $V_{max}$ in the Atlantic over the last decade (1995-2005) should terminate when the thermohaline circulation (THC) weakens in the coming decades.   Atlantic net TC frictional energy wind dissipation will then be reduced.

## *VARIATION IN MAJOR HURRICANE NUMBERS DURING THE LAST TWO DECADES OF GLOBAL WARMING*

The NCEP/NCAR reanalysis of global mean surface temperature and global SSTs shows a rise in temperature over the last 10 years in comparison with the mean global surface temperature and SST of the prior 10 years of 1985-1994.  Between 1995-2004, average global surface temperatures have been observed to be about 0.4°C higher than the earlier 10-year period of 1985-1994.  Table 1 shows the number of observed major hurricanes (Cat. 3-4-5) around the globe (excluding the Atlantic) between these two 10-year periods.  Major hurricane activity has not gone up in the more recent 10-year period when global surface temperatures have been higher.  Even when the large increase in Atlantic major hurricane activity is added to the non-Atlantic total of major hurricanes, there is no significant difference (232 vs. 256) in the numbers of global major hurricanes between these two 10-year periods.

*Table 1. Comparison of observed major (Cat. 3-4-5) tropical cyclones in all global basins (except the Atlantic) in the two most recent 10-year periods of 1985-94 and 1995-2004.*

|  | **1985-1994 (10 years)** | **1995-2004 (10 years)** |
|---|---|---|
| North & South Indian Ocean | 45 | 50 |
| South Pacific & Australia | 44 | 41 |
| NW Pacific | 88 | 87 |
| Northeast Pacific | 41 | 40 |
| GLOBE TOTAL (excluding Atlantic) | 218 | 218 |

***COMPARISON OF PACIFIC CATEGORY 4-5 TROPICAL CYCLONE ACTIVITY DURING THE LAST TWO 10-YEAR PERIODS***

The most reliable comparison of Category 4-5 hurricanes that can likely be made is to compare the last ten years (1995-2004) with the prior ten years (1985-1994) for the storm areas monitored by US and Japanese satellites. The reliable observations that are available from these two US-monitored basins in the Pacific do not indicate that the global number of tropical cyclones of Category 4-5 intensity have increased in the last 10 years when global surface temperatures have risen. Despite the last ten years of warmer global surface temperatures, the number of Category 4-5 hurricanes in the Pacific shows no increase (Table 2) and in the Atlantic, there have been prior periods in which Category 4-5 hurricanes have been as frequent as during recent years.

*Table 2. Comparison of the number of Category 4-5 tropical cyclones in the North Pacific during the last two 10-year periods.*

|  | **1985-1994** | **1995-2004** |
|---|---|---|
| NE PACIFIC | 31 | 30 |
| NW PACIFIC | 70 | 65 |
| *TOTAL* | *101* | *95* |

## *COMPARISON OF ATLANTIC HURRICANE ACTIVITY BETWEEN THE LAST 15-YEAR ACTIVE PERIOD (1990-2004) WITH AN EARLIER ACTIVE 15-YEAR PERIOD (1950-1964)*

There have been hurricane periods in the Atlantic in the past which have been just as active as the current period. A comparison of the last 15 years of hurricane activity with an earlier 15-year period from 1950-64 shows no significant difference in major hurricanes (Table 3). The number of weak tropical storms rose by over 50 percent, however. This is a reflection of the availability of the satellite in the later period. It would not have been possible that a hurricane, particularly a major hurricane, escaped detection in the earlier period. But many weaker systems far out in the Atlantic undoubtedly went undetected before satellite observations. The maximum sustained winds from 1950-1964 have been adjusted downward using the Landsea (1993) adjustment factor.

*Table 3. Comparison of Atlantic tropical cyclones of various intensities between 1950-1964 and the recent 15-year period of 1990-2004.*

|  | Cat. 4-5 | Cat. 3 | Net IH | Net H | Cat. 1-2 | TS | NS | July-August SST 10-25°N; 30-70°W |
|---|---|---|---|---|---|---|---|---|
| 1950-64 (15 yrs) | 24 | 23 | 47 | 98 | 51 | 50 | 148 | 25.69 |
| 1990-04 (15 yrs) | 25 | 18 | 43 | 100 | 57 | 78 | 178 | 26.11 |
| 1990-04 *minus* 1950-64 | +1 | -5 | -4 | +2 | +6 | +28 | +30 | +0.42 |
| Percent Increase | +4% | -22% | -9% | +2% | +12% | +56% | +18% | --- |

It should also be noted that Emanuel's large increases in net rate of TC energy dissipation (his PDI) was derived from only two of the seven global tropical cyclone basins. Data from the Northeast Pacific and the other tropical cyclone basins (North Indian Ocean and the Southern Hemisphere's three TC basins) were not a part of his study. As Table 1 indicates, these other tropical cyclone basins do not show major hurricane increases over the last two 10-year periods.

## *SUMMARY*

Despite what many in the atmospheric modeling and forecast communities may believe, there is no physical basis for assuming that global tropical cyclone intensity or frequency is necessarily related to global mean surface temperature changes of less than ±0.5$^o$C. As the ocean surface warms, global upper air temperatures also increase to maintain conditionally unstable lapse-rates and global rainfall rates at their required values. Seasonal and monthly variations within individual storm basins show only very low correlations of sea surface temperature (SST) with monthly, seasonal, and yearly variations of hurricane activity. These correlations are typically of the order of about 0.3, explaining only about 10 percent of the variance. Other factors such as tropospheric vertical wind shear, surface pressure, low level vorticity, mid-level moisture, etc. play more dominant roles in explaining hurricane variability on these shorter time scales. Although there has been a general global warming over the last 10 years, the SST increases in the individual tropical cyclone basins have been smaller (about half), and according to the observations have not brought about any significant increases in global major tropical cyclones except for the Atlantic where a shift has been made to a positive phase of the Atlantic multi-decadal oscillation. No credible observational evidence is available or likely will be available in the next few decades which will be able to directly associate global temperature change to changes in global tropical cyclone frequency and intensity.

***Acknowledgement***: I would like to acknowledge beneficial discussions on this topic with John Knaff and Philip Klotzbach.

## REFERENCES

Atkinson, G.D. and C.R. Holliday, 1977: Tropical cyclone minimum sea level pressure/maximum sustained wind relationship for the western North Pacific. Mon. Wea. Rev., **105**, 421-427.

Cocks, S.B. and W.M. Gray, 2002: Variability of outer wind profiles of western North Pacific typhoons. Classifications and techniques for analysis and forecasting. Mon. Wea. Rev., **122**, 1989-2005.

Dvorak, V.F., 1975: Tropical cyclone intensity analysis and forecasting from satellite imagery. Mon. Wea. Rev., **103**, 420-430.

Dvorak, V.F., 1984: Tropical cyclone intensity analysis using satellite data. NOAA Technical Report NESDIS 11, 45 pp.

Franklin, J.L., M.L. Black, and K. Valde, 2003: GPS dropwindsonde wind profiles in hurricanes and their operation implications. Weather and Forecasting, **18**, 32-44.

Gray, W.M., 2005: Comments on Webster, *et al.*, Science, **309**, 1844-1846.

Gray, *et al.*, 1991: Assessment of the role of aircraft reconnaissance on tropical cyclone analysis and forecasting. BAMS, **72**, 1867-1883.

Knaff, J.A. and R.M. Zehr, 2005: Reexamination of tropical cyclone pressure wind relationships (being submitted for publication).

Landsea, C. W., 1993: A climatology of intense (or major) Atlantic hurricanes. Mon. Wea. Rev.*,* **121**, 1703–1713.

Martin, J.D. and W.M. Gray, 1993: Tropical cyclone observation and forecasting with and without aircraft reconnaissance. Weather and Forecasting, **8**, 519-532.

Merrill, R.T., 1984: A comparison of large and small tropical cyclones. Mon. Wea. Rev., **112**, 1408-1418.

Weatherford, C.L. and W.M. Gray, 1988a: Typhoon structure as revealed by aircraft reconnaissance. Part I: Data analysis and climatology. Mon. Wea. Rev., **116**, 1032-1043.

Weatherford, C.L. and W.M. Gray, 1988b: Typhoon structure as revealed by aircraft reconnaissance. Part II: Data analysis and climatology. <u>Mon. Wea. Rev.</u>, **116**, 1044-1056.

Webster, P.J., et al., 2005: Changes in tropical cyclone number, duration, and intensity in a warming environment. <u>Science</u>, **309**, 1844-1846.